%% file: aanda_v3_arxiv.tex
\documentclass{aa}  

\usepackage{graphicx}
\usepackage{txfonts}

\usepackage[pdfpagelabels=false]{hyperref}	
\hypersetup{colorlinks=true,linkcolor=blue,citecolor=blue,filecolor=blue,urlcolor=blue,}

\begin{document} 

   \title{Spectral Stacking of Radio-Interferometric Data}


   \author{Lukas Neumann
          \inst{1}\fnmsep\thanks{\email{lukas.neumann.astro@gmail.com}}
          \and
          Jakob S. den Brok\inst{1,2}
          \and
          Frank Bigiel\inst{1}
          \and
          Adam Leroy\inst{3}
          \and
          Antonio Usero\inst{4}
          \and
          Ashley~T.~Barnes\inst{5}
          \and
          Ivana~Be\v{s}li\'c\inst{1,6}
          \and
          Cosima Eibensteiner\inst{1}
          \and
          Malena Held\inst{1}
          \and
          Mar\'ia J. Jim\'enez-Donaire\inst{4}
          \and
          Jérôme~Pety\inst{6,7}
          \and
          Erik~W.~Rosolowsky\inst{8}
          \and
          Eva Schinnerer\inst{9}
          \and
          Thomas G.~Williams\inst{10}
          }

   \institute{
    Argelander-Institut für Astronomie, Universität Bonn, Auf dem Hügel 71, 53121 Bonn, Germany
    \and 
    Center for Astrophysics $\mid$ Harvard \& Smithsonian, 60 Garden St., 02138 Cambridge, MA, USA
    \and    
    Department of Astronomy, The Ohio State University, 140 West 18th Ave, Columbus, OH 43210, USA
    \and    
    Observatorio Astron\'omico Nacional (IGN), C/ Alfonso XII, 3, E-28014 Madrid, Spain
    \and
    European Southern Observatory, Karl-Schwarzschild Stra{\ss}e 2, D-85748 Garching bei M\"{u}nchen, Germany
    \and
    LERMA, Observatoire de Paris, PSL Research University, CNRS, Sorbonne Universit\'es, 75014 Paris, France
    \and
    Institut de Radioastronomie Millim\'etrique (IRAM), 300 Rue de la Piscine, F-38406 Saint Martin d’Hères, France
    \and
    Dept. of Physics, University of Alberta, Edmonton, Alberta, T6G 2E1, Canada
    \and
    Max Planck Institute for Astronomy, K\"onigstuhl 17, D-69117 Heidelberg, Germany
    \and
    Sub-department of Astrophysics, Department of Physics, University of Oxford, Keble Road, Oxford OX1 3RH, UK
    }

   \date{Received 13 February 2023; accepted 28 April 2023}

  \abstract
   {Mapping molecular line emission beyond the bright low-$J$ CO transitions is still challenging in extragalactic studies, even with the latest generation of (sub-)mm interferometers, such as ALMA and NOEMA.}
   {We summarise and test a spectral stacking method that has been used in the literature to recover low-intensity molecular line emission, such as HCN(1--0), HCO$^+$(1--0), and even fainter lines in external galaxies. The goal is to study the capabilities and limitations of the stacking technique when applied to imaged interferometric observations.}
   {The core idea of spectral stacking is to align spectra of the low S/N spectral lines to a known velocity field calculated from a higher S/N line expected to share the kinematics of the fainter line, e.g., CO(1--0) or 21-cm emission. Then these aligned spectra can be coherently averaged to produce potentially high S/N spectral stacks. Here, we use imaged simulated interferometric and total power observations at different signal-to-noise levels, based on real CO observations.}
   {For the combined interferometric and total power data, we find that the spectral stacking technique is capable of recovering the integrated intensities even at low S/N levels across most of the region where the high S/N prior is detected. However, when stacking interferometer-only data for low S/N emission, the stacks can miss up to 50\% of the emission from the fainter line.}
   {A key result of this analysis is that the spectral stacking method is able to recover the true mean line intensities in low S/N cubes and to accurately measure the statistical significance of the recovered lines. To facilitate the application of this technique we provide a public Python package, called \textsc{PyStacker}.}

   \keywords{Methods: data analysis --
             Techniques: interferometric --
             Galaxies: ISM --
             Radio lines: galaxies --
             Radio lines: ISM
            }

   \maketitle
%
\begin{figure*}
    \centering
    \includegraphics[width=\textwidth]{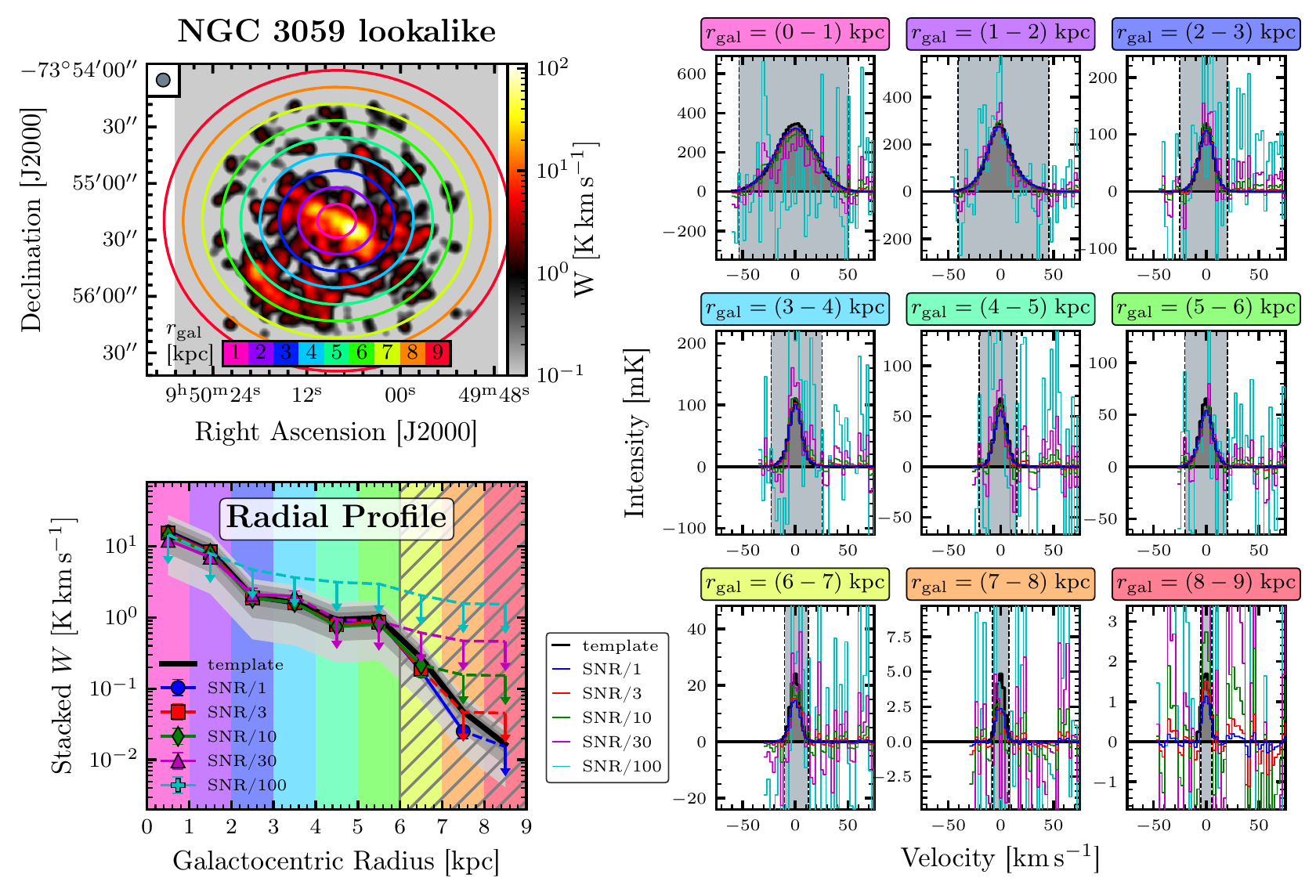}
    \caption{Spectral stacking of the NGC\,3059 template galaxy via galactocentric radius. \textit{Top left:} Moment-0 map of the true, i.e. known, simulated CO(2--1) emission mimicking the intensity distribution of NGC\,3059. The data have been convolved to the common highest resolution over all array configurations, i.e. $7.2''$, indicated by the grey circle in the upper left. The coloured rings show loci of galactocentric radius. \textit{Right panels:} Stacked spectra in 1\,kpc radial bins from the centre out to 9\,kpc as illustrated in the top left panel. The dark grey histogram shows the true spectra, i.e. the spectra obtained by stacking the input template. The coloured lines show the stacked spectra from the respective simulated data cubes. The grey-shaded area indicates the velocity range used to compute the integrated intensities. \textit{Bottom left:} Stacked integrated intensities corresponding to the spectra shown in the right panels plotted against galactocentric radius. The black line shows the true radial trend, and the coloured lines show the recovered trend of the simulated data cubes. Solid points indicate data above $3\sigma$; downward pointing arrows denote $3\sigma$ upper limits. The grey shaded areas show differences to the true trend in levels of $\pm\{25,50,75\}\%$. The hatched area denotes the regime, where the prior, i.e. SNR/1, is detected ($\mathrm{S/N}\geq 3$) in less than 20\% of the pixels.}
    \label{fig:ngc3059_stacking}
\end{figure*}

\section{Introduction}
Mapping extragalactic molecular line emission with high spatial resolution and sensitivity is still challenging even with the latest generation of (sub-)mm interferometers, such as ALMA (Atacama Large Millimeter/submillimeter Array) and NOEMA (Northern Extended Millimeter Array). 
Practically, for most nearby galaxies, only the low-$J$ CO transitions, which are the brightest mm-wave lines, can be rapidly surveyed at good $\lesssim 1''$ resolution while also achieving widespread high significance detections across the full disc of a typical star-forming galaxy \citep[e.g.][]{Leroy2021b}.
Recovering integrated intensities of fainter, and hence typically low signal-to-noise ratio (S/N) lines, like HCN(1--0), HCO$^+$(1--0) or N$_2$H$^+$(1--0), is more challenging.
These lines carry critical physical information on the composition, temperature, and density of the gas but often have intensities 30 to $>$100 times fainter than the CO lines \citep[e.g.,][]{Usero2015, Jimenez-Donaire2017}.
To measure the intensities of these other lines ``spectral stacking'' methods have become popular in recent years.

Stacking of astronomical data has been used for at least four decades \citep[e.g.,][]{Cady1980} and applied across wavelength regimes, from X-ray \citep[e.g.][]{Hickox2007, Chen2013} to sub-millimetre and radio wavelengths \citep[e.g.][]{Knudsen2005, Karim2011, Schruba2011, Delhaize2013, Caldu-Primo2013, Bigiel2016, Lindroos2016, Jolly2020}. In the past decade, spectral stacking has become a particularly important tool in millimetre studies of galaxies, allowing the recovery of otherwise undetected line emission. 
For example, \citet{Jimenez-Donaire2019}, \citet{Beslic2021} and \citet{Neumann2023} all use spectral stacking leveraging a CO emission prior to recover emission from faint, high critical density emission lines, including HCN(1--0), HNC(1--0) or HCO$^+$(1--0), across large areas in the disks of nearby galaxies. 
\citet{denBrok2021}, \citet{denBrok2022} used spectral stacking based on $^{12}$CO to obtain more significant constraints on lines tracing rarer CO isotopologues. 
And \citet{Schruba2011} used 21-cm emission as a prior to construct extended, sensitive radial profiles of CO emission even in the outer parts of galaxies. 
These studies all demonstrate how spectral stacking recovers more information about the distribution, composition, and physical conditions of the molecular gas in galaxies.

The basic idea of spectral stacking as often applied to nearby galaxies is to align all spectra by recentering them on the local mean velocity of the interstellar medium (ISM), which is measured using a high S/N prior, e.g., CO(1--0) or the 21-cm line (Sec.~\ref{sec:stacking}). 
Then spectra from different parts of the galaxy can be coherently averaged with minimal contributions from noise in empty parts of the bandpass. By averaging in azimuthal rings one can construct sensitive radial profiles.
One can also average as a function of other quantities to test specific hypotheses or scaling relations, e.g. galactocentric radius, line intensity, surface density, or star formation rate. 
Carrying out this stacking on the spectra allows an important visual check that the averaged result indeed looks like an astrophysical spectral line, i.e. to first order a Doppler-broadened Gaussian line profile, and can even allow recovery of mean kinematic information via the width of the Gaussian.

While these techniques are simple in principle, a key uncertainty remains surrounding their application to the most powerful current mm-wave telescopes, ALMA and NOEMA. 
These facilities are interferometers and the images they produce reflect both incomplete sampling of the $u-v$ plane and a deconvolution process that often focuses on bright emission. 
While $u-v$ plane stacking can alleviate both concerns in unresolved objects or those with simple geometries, stacking in the image plane remains the most practical option for extended, complex sources like nearby galaxies. 
Since stacking using these powerful telescopes represents a key way to push our knowledge of the physical state and makeup of the ISM, evaluating the accuracy of this technique when applied to recover faint, low S/N lines from interferometer data is a key next step.

The goal of this work is to provide such a demonstration. For this purpose, we use the CASA \citep[Common Astronomy Software Applications, ][]{CASA2022} ALMA simulator to simulate interferometric and total power observations of low S/N lines based on a known input model. 
The resulting simulated observations are imaged using the PHANGS (Physics at High Angular resolution in Nearby GalaxieS)--ALMA pipeline \citep{Leroy2021a}.
Then we apply the spectral stacking method and assess how well the stacks recover the known input. In particular, we stack via the galactocentric radius using simulated CO(2--1) data cubes built on real observations of galaxies from the PHANGS--ALMA survey \citep{Leroy2021b}.

We also present a new public python package, \textsc{PyStacker} that can be used to easily apply these techniques. 
This utility complements the tool \textsc{LineStacker} presented by \citet{Jolly2020}, which is also validated against simulation. 
Their work focuses on spectrally stacking many distinct sources in three dimensions, while our code emphasizes stacking within an individual data set in the presence of a complex prior velocity field.
Another stacking package called \textsc{spectral-stack}\footnote{\url{https://github.com/low-sky/spectral_stack}}, relies on Fourier shifting to align the spectra to be averaged. The advantage of this approach is that the noise properties and channel-to-channel correlations are preserved. However, it deals less well with edge effects.

\section{Description of the Spectral Stacking Method} 
\label{sec:stacking}

\begin{figure}
    \centering
    \includegraphics[width=\columnwidth]{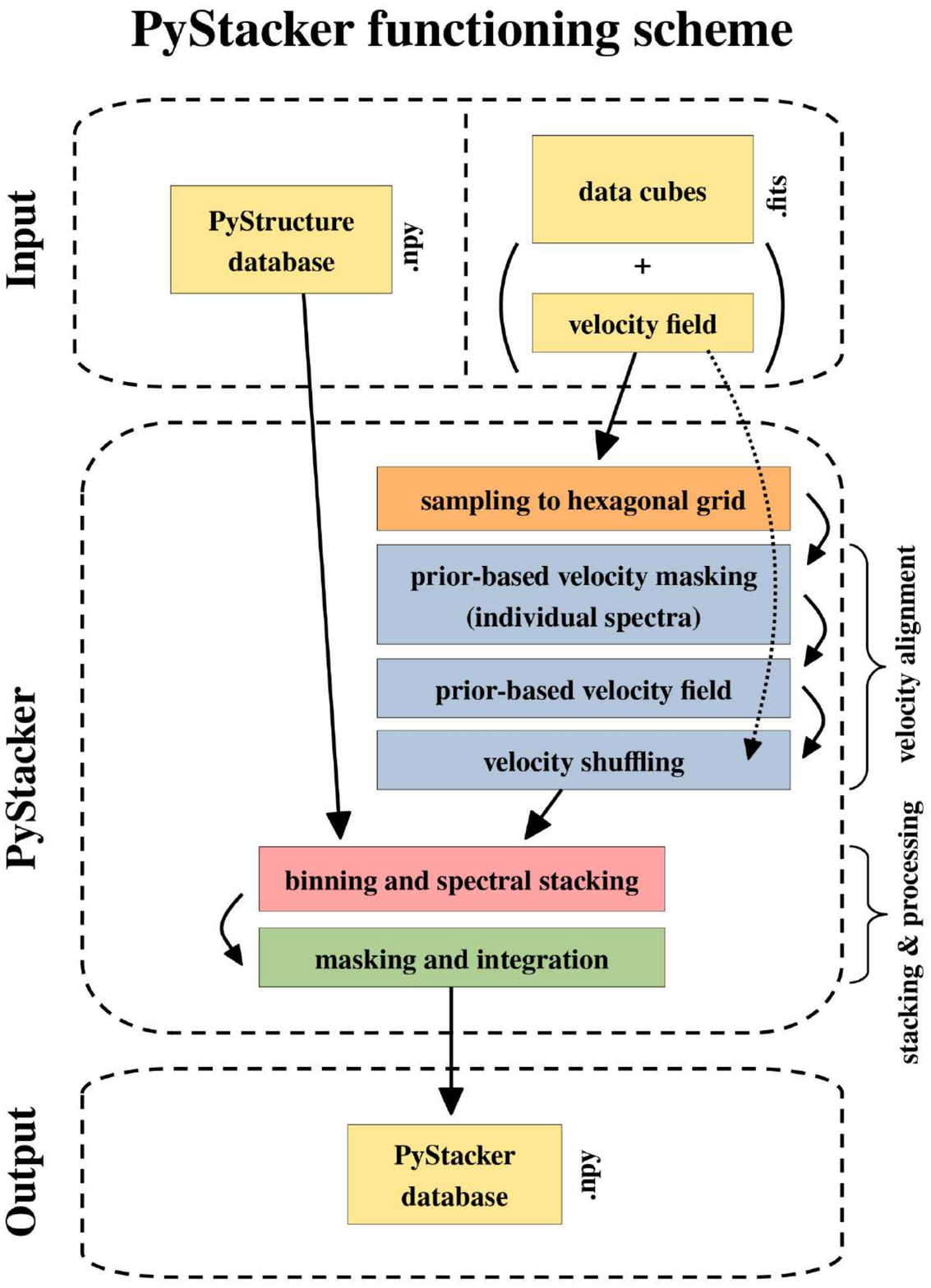}
     \caption{Schematic of the functioning principle of the \texttt{PyStacker} package. There are two possible input option. One is the ``PyStructure'' database, which is a numpy dictionary containing all the molecular line emission data. The PyStructure database is produced by a separate pipeline and already contains the velocity alignment analogous to the velocity shuffling performed by \texttt{PyStacker}. Second, and the default usage for most users, the input can be data cubes in the form of \texttt{.fits} files, where each FITS file contains the position-position-velocity information of the respective spectral line. Here, the user can provide a model velocity field used to shuffle the velocity field of the lines to be stacked. In addition, in both cases, a configuration file must be specified, which sets the parameters for the stacking. If data cubes are provided, those are sampled on hexagonal gridded pixels with half-beam spacing. Next, the significant pixels of the given prior are identified and the velocity of the peak intensity of each prior-detected spectrum is used to define the velocity field (if not provided by the input model). After applying the velocity offsets to the molecular line data, the spectra inside the given bins are averaged. The user can specify in the configuration file if the prior-non-detected pixels are ignored or set to zero for the bin average. Afterwards, the stacked spectra of the prior are used to build a velocity mask for each stack, which is used to compute the integrated intensities (inside the mask) and the uncertainties (rms outside the mask). The final output is a numpy dictionary containing the stacked spectra along with their integrated intensities, uncertainties and other quantities (see the documentation for the full output content).}
    \label{fig:pystacker_diagram}
\end{figure}

The main goal of the spectral stacking technique is the recovery of low S/N lines by shifting the spectra to a known velocity field (defined by a high S/N prior, e.g. CO(1--0) or CO(2--1) or the \textsc{Hi} 21-cm line) and then averaging the spectra based on another parameter such as environment, star formation rate or line intensity.
Our stacking method is based on \citet{denBrok2022} and \citet{Neumann2023} and our implementation is available as a python package, called \textsc{PyStacker}\footnote{\url{https://github.com/PhangsTeam/PyStacker}}.
We will describe the basic steps of the code in the following.

We begin with a set of data cubes of the same target with different S/N levels of the input line emission. First, we homogenise the data bringing all data cubes to the same coordinate grid and convolving to the same spatial resolution.
Then, we define a prior, typically the most significantly detected line, which is used to obtain the velocity field as the velocity at the peak intensity of each spectrum\footnote{Note that the code also allows inputting a model velocity field.} \citep[][]{Koch2018RNAAS}.
We use this velocity field to redefine the spectral axis for each individual spectrum in the cube\footnote{Recentering the spectrum itself has some associated subtleties, and can be done using either Fourier techniques or via regridding and oversampling. In this paper, we use re-gridding techniques but note that the choice of approach can affect the channel-to-channel correlation and noise properties of the stacked spectrum.}
so that the emission of all lines should be centred at a velocity of 0$\,$km$\,$s$^{-1}$.
The result is sometimes referred to as a ``shuffled'' cube in reference to the GIPSY \citep[Groningen Image Processing System;][]{vanderHulst1992} task \texttt{shuffle}.

One can then average the shuffled spectra inside bins defined by any arbitrary quantity of scientific interest, e.g. galactocentric radius, CO(1--0) line intensity, or star formation rate. For instance, in this work, we stack as a function of galactocentric radius (Sec.~\ref{sec:W_recovery}). 
If signal is present in the stacked line within a given bin, the averaged spectrum should then appear as a clear emission line (e.g., Fig.~\ref{fig:ngc3059_stacking} right panel). 
For comparison, averaging across different parts of strongly rotating discs without first adjusting to the local velocity yields a broad, lower signal-to-noise profile (e.g., see Fig. 2 in \citealt{Schruba2011}).

We can only reliably ``shuffle'' the spectra if the prior is actually detected in the respective spectrum. Thus, if the prior is only detected in a fraction of the spectra inside a bin, we rely on these spectra to infer the average of the bin.
In this case, we consider the emission in the spectra that could not be shuffled to be equal to zero, such that the average is always measured relative to all spectra inside a given bin\footnote{The alternative, i.e. averaging over the prior-detected spectra only, tends to overestimate the stacked intensity. However, \texttt{PyStacker} allows us to also use this option and we show its results in Fig.~\ref{fig:ngc3059_ignore_empties}.}. 
We expect this to be a reasonable assumption when stacking a rare, faint molecular line like HCN(1--0) using CO as a prior\footnote{In other applications of this method, a lower resolution cube or even a model rotation curve may sometimes be used as a prior to shuffle in cases where the brighter line has patchy coverage or limited S/N.}.

Within any given bin $n$, we measure the average spectrum $T_{n,\mathrm{stack}}(v)$:
\begin{align}
    T_{n,\mathrm{stack}}(v) = \dfrac{1}{N_{\rm tot}(n)} \sum\limits_{i=0}^{N_{\rm det}(n)} T_{n,i}(v) \;,
    \label{equ:stacked_spectrum}
\end{align}
where $N_{\rm tot}(n)$ and $N_{\rm det}(n)$ are the total number and the prior-detected number of spectra in bin $n$.

To compute the integrated intensities of the stacked spectra, we build a mask based on the high-S/N reference cube. We select the velocity range of significant emission for each spectrum as described in \citet{Beslic2021} and integrate the intensities over mask-selected velocity channels.
\begin{align}
    W_{n} = \sum\limits_{N_{\rm mask}} T_{n,\mathrm{stack}}(v) \cdot \Delta v_{\rm channel} \;,
    \label{equ:stacked_W}
\end{align}
where $\Delta v_{\rm channel}$ is the channel width and $N_{\rm mask}$ is the number of (independent) channels inside the mask.

The nominal uncertainties of the integrated intensities ($\sigma_W$, studied in Sec.~\ref{sec:uncertainties}) are given by:
\begin{equation}
    \sigma_W = \mathrm{rms} \times \sqrt{\dfrac{N_\mathrm{tot}}{N_\mathrm{det}}} \times \Delta v_{\rm channel} \times \sqrt{N_{\rm mask}} \,,
    \label{equ:unc_measured}
\end{equation}
where rms is the root mean square of the emission-free channels, i.e. outside the mask, in the stacked spectrum.
Since the stacked spectrum is computed from the prior-detected pixels, $N_\mathrm{det}$, but divided by the total number of pixels in that bin, $N_\mathrm{tot}$, (Eq.~\ref{equ:stacked_spectrum}) the measured rms of the emission-free channels is biased low if $N_\mathrm{det}<N_\mathrm{tot}$. Therefore, we have to correct the rms by the factor $\mathbf{\sqrt{N_\mathrm{tot}/N_\mathrm{det}}}$ ($\geq 1$) in Eq.~\ref{equ:unc_measured} in order not to underestimate the rms and thus $\sigma_W$. The correction factor mimics the increase in noise when adding up $N_\mathrm{tot}$ spectra with the same noise level.

\section{Recovery of Integrated Intensities} 
\label{sec:W_recovery}

\begin{figure}
    \centering
    \includegraphics[width=\columnwidth]{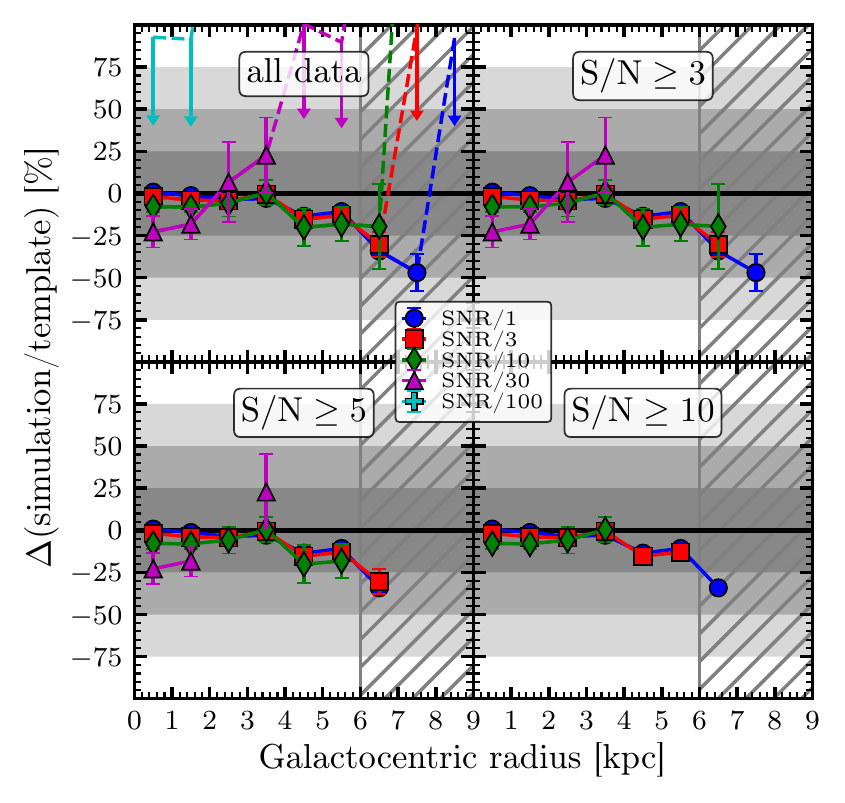}
     \caption{Agreement between measured and expected radial trends at different sigma clipping levels. Shown is the relative differences between the measured radial stacks from the simulated 12m+7m+tp cubes and the true stacks as follows from the bottom left panel of Fig.~\ref{fig:ngc3059_stacking} by subtracting the template trend. The difference between the different panels is that we clip data of the resulting stacks at 3, 5, or 10 S/N ($W\sigma_W$). Note that the stacking procedure is always the same. The hatched area denotes the regime, where the prior, i.e. SNR/1, is detected in less than 20\% of the pixels.}
    \label{fig:ngc3059_sigma_clipping}
\end{figure}

We apply our method to simulated data cubes with known input to test how well spectral stacking can recover the integrated intensities of molecular line emission as a function of the noise level of the observations. 

Specifically, we use a set of simulations of molecular line emission produced to validate the PHANGS--ALMA data reduction pipeline \citep{Leroy2021a}. As described in \citet{Leroy2021a}, the simulated CO(2--1) data cubes have been produced using the CASA tasks \texttt{simdata} and \texttt{simobserve} using inputs based on real PHANGS--ALMA CO(2--1) images. The simulated observations mimic interferometric observations of the galaxy NGC\,3059 similar to the PHANGS--ALMA survey\footnote{Note that the NGC\,3059 lookalike has been rotated such that the major axis of the galaxy is aligned with the declination. However, these modifications have no effect on our analysis.} The simulations included the creation of a simulated total power map constructed by convolving the input model to the resolution of the ALMA TP antennas and adding Gaussian noise of the expected magnitude for a real PHANGS--ALMA TP observation.

The input intensity cube, hereafter referred to as ``template'', is the actual masked NGC\,3059 cube from PHANGS--ALMA. We show the integrated intensity map of this template data cube in the upper left panel of Fig.~\ref{fig:ngc3059_stacking}. These ``true'' data are used to construct simulated  12m, 7m, and total power observations and imaged via the PHANGS--ALMA pipeline \citep[for more details see][]{Leroy2021a}. Then we run these through the stacking pipeline in this study. The use of real data as a model means that there will be some observational noise in the true data, but we consider that as signal and explore how well it gets recovered, and it should have only a modest impact on the analysis.

The simulations produce images for different combinations of the ALMA main array, the ACA 7-m antennas, and the total power data, i.e. 12m+7m+tp, 12m+7m, 7m+tp, 12m and 7m (Sec.~\ref{sec:arrays}). They also produce cubes with a range of different signal-to-noise levels, which we refer to as ``SNR/1'', ``SNR/3'', and so on. The ``SNR/1'' cube mimics the sensitivity of a typical PHANGS--ALMA CO(2--1) observation, while ``SNR/3'', ``SNR/10'', ``SNR/30'', ``SNR/100'' have a factor of 3, 10, 30, 100, respectively, lower S/N\footnote{Note that the original exercise in \citet{Leroy2021a} actually scales the signal down by factors of 3, 10, 30, 100} but leaves the noise the same. Here, we take these cubes and rescale them with the respective factors to obtain cubes at the same intensity but different noise and thus S/N levels. At the common PHANGS--ALMA sensitivity and for a brightness distribution similar to NGC\,3059, SNR/10 could be representative for $^{13}$CO(1--0), SNR/30 for HCN(1--0) or HCO$^+$ and SNR/100 for fainter lines like N$_2$H$^+$(1--0).

The S/N levels of the moment-0 maps resulting from the various S/N cubes range from 2.5 (minimum), 341.7 (maximum) for the SNR/1 data, over -1.9 (minimum), 32.5 (maximum) for the SNR/10 cube, down to -3.8 (minimum), 3.7 (maximum) for the version with 100 times higher noise.
This means we study cubes that contain significant emission across most of the field of view all the way to almost pure noise cubes.

The angular resolution of the 12m+7m+tp cube is 2.7$^{\prime\prime}$ and higher than that of the 7m and 7m+tp cubes at 7.2$^{\prime\prime}$, which correspond to a linear scales of 264~pc and 702~pc, respectively, at a distance of 20.2\,Mpc. 
In order to compare the 12m+7m+tp results more directly with the other arrays we convolve all cubes to a common 7.2$^{\prime\prime}$ resolution and focus on the 12m+7m+tp at 7.2$^{\prime\prime}$ resolution for most of the analysis.
We note that the convolution from the native 12m resolution (2.7$^{\prime\prime}$) to the common best resolution (7.2$^{\prime\prime}$) might smear out some of the significant, compact emission and thus potentially reduce the efficiency of the stacking.
However, we checked that, in our case, the recovered stacks from the native resolution cubes are consistent with the stacks from the convolved cubes.

We apply the spectral stacking method described in Sec.~\ref{sec:stacking} to the five data cubes at different S/N levels described above. 
We stack the spectra by galactocentric radius from 0 to 9\,kpc in 1\,kpc increments as illustrated in Fig.~\ref{fig:ngc3059_stacking}.
We note that there is very little emission, i.e. less than 20\% of the pixels in the SNR/1 moment-0 map contain significant emission, outside of 6\,kpc.
Therefore, we limit most of the discussion to the inner 6\,kpc and consider this the typical extent of the molecular gas disc.
For the outer radii, an alternative approach could be to average over a larger region, e.g. bin everything beyond the radius of 6\,kpc, in order to potentially recover more of the fainter emission at the cost of spatial information. However, in this case, we do not recover more emission due to the steep drop in emission beyond 6\,kpc.
We use the SNR/1 data (i.e. the simulated data mimicking PHANGS--ALMA CO(2--1) observations) as a prior to account for the varying velocity field across the galaxy and to determine the channels of significant emission to compute the integrated intensities.
For each configuration, the respective SNR/1 cube is used as a prior. This means, for stacking of 12m+7m+tp cubes we use the 12m+7m+tp SNR/1 cube as the prior, for stacking of 7m+tp data the 7m+tp SNR/1 cube, etc. This approach is similar to how we typically handle real observational data, where different lines have been observed with the same interferometric setup.
In this case, by construction, the velocity fields of the different line cubes are identical. In reality, we may expect small velocity offsets between different spectral lines leading to slightly broader and thus potentially less significant stacked lines. Though, this effect is expected to be small when studying various molecular lines, which should share similar kinematics.
The stacked spectra are shown in the right panels of Fig.~\ref{fig:ngc3059_stacking}.
Since, in this case, we know the true velocity distribution from the template cube, we repeat the same procedure using the true velocity field (Appendix~\ref{sec:appendix}).

The resulting radial profiles of the stacked integrated intensities are shown in the bottom left panel of Fig.~\ref{fig:ngc3059_stacking}.
Overall, we find that the radial trend is well recovered across most of the molecular disc down to the SNR/10 data cubes.
In the SNR/30 cube, we are still able to recover the radial trend out to 4~kpc, where, in the 3 to 4~kpc bin, the median moment-0 S/N is 0.54.
For the noisiest data used here (SNR/100) we obtain only upper limits which highlights the fact that it is extremely challenging to map molecular discs of nearby galaxies in line emission that is $\sim$~100 times fainter than CO(2--1), e.g. the popular Galactic dense gas tracer N$_2$H$^+$. 
PHANGS--ALMA integrated for $\sim 1$ minute per field. 
This exercise implies that to achieve the S/N$\sim 10$ required for reliable stacked detections, integrations $\sim 100$ times longer, i.e., $\sim 2$~hours per pointing, would be required.
Another approach could be to modify the binning, e.g. by averaging spectra over larger regions. While losing spatial information and was not performed here, this could potentially recover otherwise undetected emission. Therefore, we recommend adapting the binning parameters to the strength and distribution of the studied line emission.
We do emphasise that NGC~3059 is a relatively low luminosity galaxy, and the situation may be more optimistic in somewhat brighter targets.

We highlight the differences between the recovered stacks and the true values in Fig.~\ref{fig:ngc3059_sigma_clipping}.
Based on the computed uncertainties of the stacked integrated intensities, $\sigma_w$, we clip at S/N ($W/\sigma_W$) levels of 3, 5 and 10.
As expected, we find that with stronger $\sigma_W$-clipping the stacked line intensities show better agreement with the true values such that for data above $10\sigma$ the maximum discrepancy is $<35\%$, and $<15\%$ ($<8\%$) in the inner 6\,kpc (4\,kpc).
Certainly, we also find systematically too low values at larger radii, i.e. apparently significant measurements that do not agree within the uncertainties with the true values for $r_{\rm gal}>4\,$kpc.
This offset might be explained by the low fraction of spectra contributing to the stacks in these bins (see Table~\ref{tab:prior_detection}).
For the outer bins, i.e. $r_{\rm gal}>4\,$kpc, less than half of the spectra inside each bin could be used for stacking and as a result, we may potentially miss some emission hidden in the noise that we are not able to recover. 
However, we find a very similar discrepancy if the velocity field is perfectly known (Fig.~\ref{fig:ngc3059_template_prior}), which suggests that the offset is at least partly arising from the imaging and not the stacking procedure.
Nevertheless, we find, over all S/N cubes, an agreement between the significant stacked line intensities and the true values within 23\% in bins where the prior is at least moderately ($\geq 36\%$ of the pixels) detected, i.e. within 6\,kpc.
These results demonstrate that the quality of the stacking results is linked to the significance of the prior used to align the velocity field and the imaging of the interferometric data.

\begin{figure}
    \centering
    \includegraphics[width=\columnwidth]{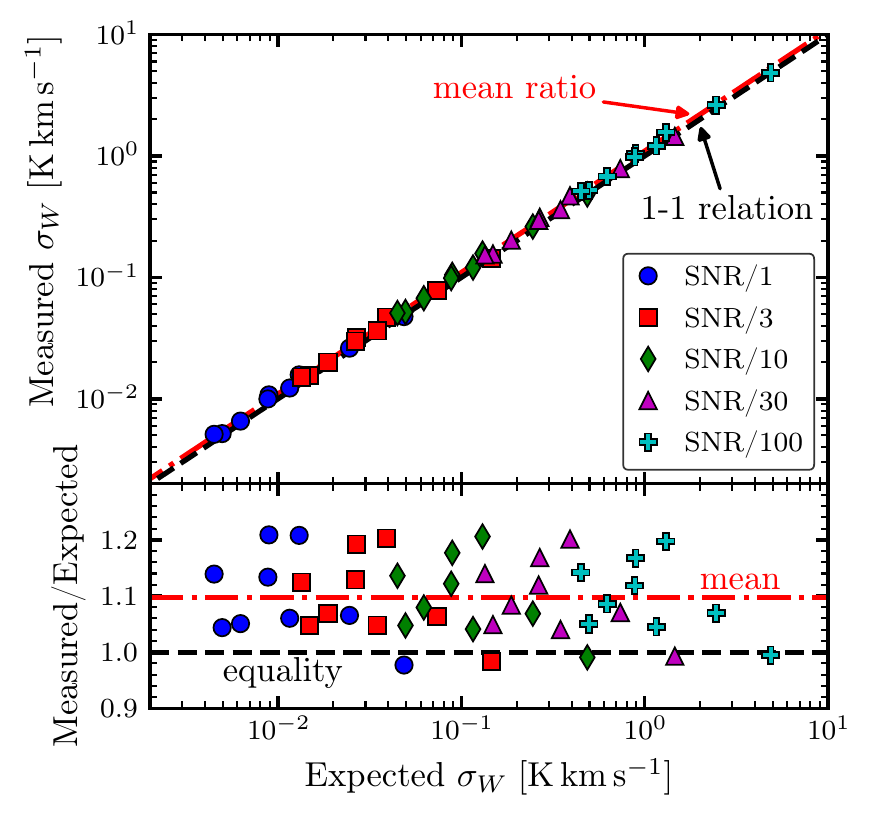}
     \caption{Measured vs. expected uncertainties. \textit{Top:} Comparison between measured (12m+7m+tp) and expected uncertainties of the integrated stacks. The measured uncertainty is obtained from the emission free channel of the stacked spectra following Eq.~\ref{equ:unc_measured}. The expected uncertainty is inferred via Gaussian error propagation from the SNR/100 cube treated as a noise cube. \textit{Bottom:} The ratio between the measured and the expected uncertainties against the expected uncertainties.}
    \label{fig:ngc3059_uncertainties}
\end{figure}

\subsection{Uncertainties} 
\label{sec:uncertainties}

For interpreting the results, it is crucial to have a robust measure of the uncertainties and the resulting S/N in order to infer if a data point is significant or not.
We measure the uncertainties of the stacked integrated intensities from the standard deviation in the emission-free channels following Eq.~\ref{equ:unc_measured}.
Here we check whether this uncertainty matches the uncertainty obtained by propagating the noise measured in the cube.

To do so, we take the SNR/100 cube, which does not contain any significant spectra and consider it as a pure noise cube.
We compute the rms in each pixel as the standard deviation across the corresponding pixel.
Next, we bin the noise map in radial increments, analogous to radial stacking, and propagate the uncertainty to obtain the expected rms of the stacked spectrum in each bin.
The propagated uncertainty is computed as the average rms in each bin corrected for the number of pixels by dividing by the square root of the number of pixels in that bin.
Finally, the expected uncertainty is computed analogously to the measured uncertainty (Eq.~\ref{equ:unc_measured}), but using the cube-propagated rms.
We re-scale the cube-propagated rms to the respective noise cubes by multiplying with the respective noise level factors and plot the measured against the expected uncertainties for all S/N cubes (Fig.~\ref{fig:ngc3059_uncertainties}).
In the Appendix, we also show the resulting S/N of the stacks, i.e. $W/\sigma_W$, and compare the measured and expected S/N (Fig.~\ref{fig:ngc3059_snr}).

We find that the measured uncertainties are strongly correlated with the expected uncertainties, but slightly biased by $\sim 10\,\%$ on average and little scatter within $\pm 10\,\%$.
The slightly too large measured uncertainties could arise from some emission remaining in the assumed emission-free channels after masking, which contributes to the rms estimation.
These results demonstrate that we measure trustworthy statistical uncertainties on the stacked integrated intensities.

\subsection{Array Configurations} 
\label{sec:arrays}

\begin{figure}
    \centering
    \includegraphics[width=\columnwidth]{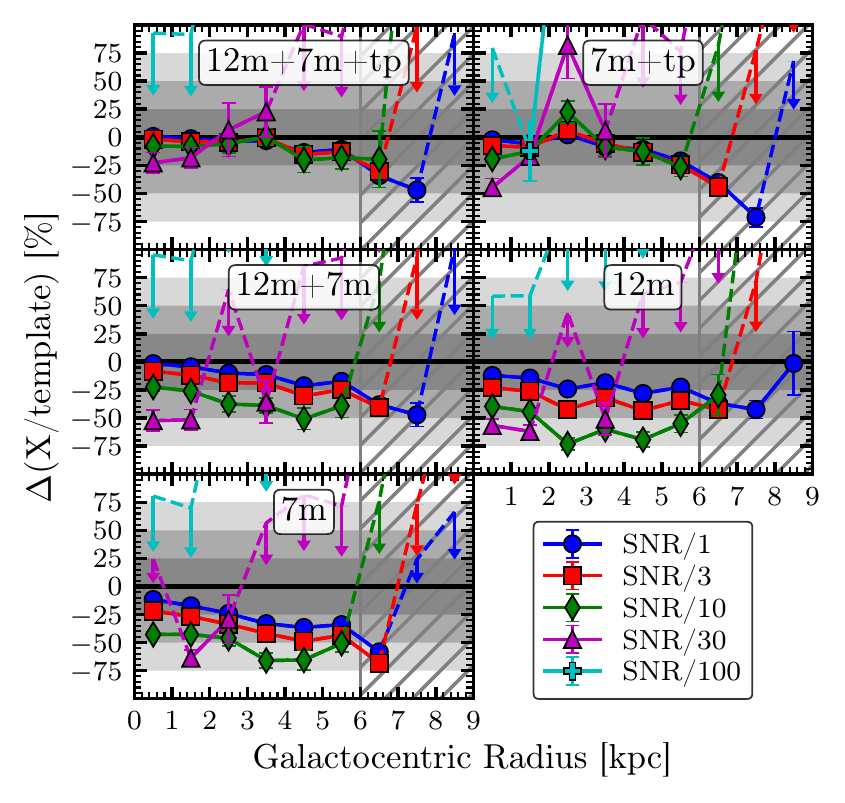}
     \caption{Flux recovery using different array configurations. Comparison between radial stacking obtained from different array configurations ($\mathrm{X}=\{\mathrm{12m+7m+tp}, \mathrm{7m+tp}, \mathrm{12m+7m}, \mathrm{12m}, \mathrm{7m}\}$), as indicated in the top of each panel. Shown is the ratio between the radial stacks obtained from the simulated data cubes at the given combination of telescope arrays and the true, template values against the galactocentric radius. Solid points show data above $3\sigma$ and downward pointing arrows denote $3\sigma$ upper limits. The hatched area denotes the regime, where the prior, i.e. SNR/1, is detected in less than 20\% of the pixels.}
    \label{fig:ngc3059_arrays}
\end{figure}

Interferometric observations filter out the extended emission of the source if not combined with single-dish data.
Using the simulated observations, we can study how well interferometric data alone can recover line emission in radial bins, and so test whether total power data are needed to obtain accurate stacking results.
We repeat the above-described spectral stacking method using data obtained from combining different telescope array configurations, i.e. ``7m+tp'' (the ACA including total power data), ``12+7m'' (the main array and ACA 7-m antennas), ``12m'' (the main array alone), and ``7m'' (the ACA 7-m data alone) \citep[see][for more information]{Leroy2021a}.

Fig.~\ref{fig:ngc3059_arrays} presents the radially stacked line intensities from the above-listed configurations relative to the template values.
The 12m+7m+tp should recover all spatial scales and can be considered the benchmark for the other configurations.
We find that the 7m+tp data performs similarly to the 12m+7m+tp, though with a significantly larger scatter which is expected due to the lower sensitivity.
For the pure interferometric data, i.e. 12+7m, 12m and 7m, we find systematically too low stacked line intensities at all radii, especially when considering 7m only, where we miss $10-20\%$ across all bins even for the SNR/1 data.
Most interestingly, we find a trend with S/N, i.e. the lower the S/N of the cube, the larger the bias of the stacks. 
In the most extreme case, i.e. 7m SNR/10, the radial profile is detected out to 6$\,$kpc, but yields $30-50\%$ lower line intensities, than obtained with the 12m+7m+tp configuration.
These results are in line with the conclusions about the spatial filtering of interferometric data drawn in \citet{Leroy2021a} and enforce the need for total power observations in order to cover the flux information from small spatial scales.

\subsection{Weighted Stacking} 
\label{sec:weighted_stacking}

\begin{figure}
    \centering
    \includegraphics[width=\columnwidth]{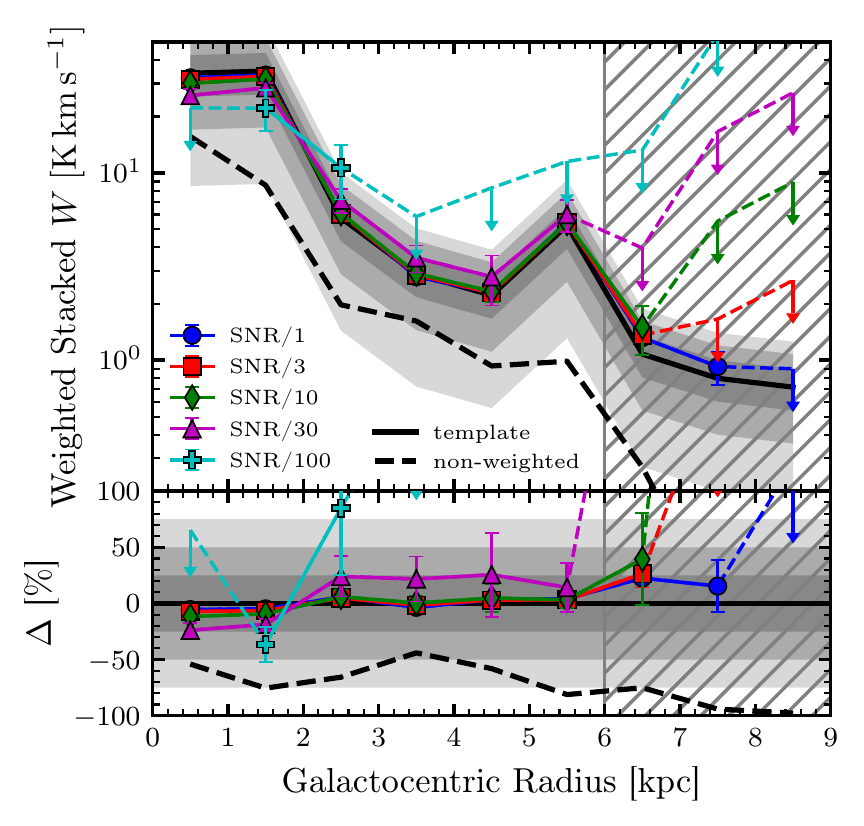}
     \caption{Intensity-weighted stacking. \textit{Top:} Radial stacking similar to Fig.~\ref{fig:ngc3059_stacking}, but using the intensity of the prior, i.e. the ``SNR/1'' integrated intensities, as a weight following Eq.~\ref{EQU:weighted_stacking}. The black solid line shows the template, i.e. the true, radial profile, where the template line intensities are used as the weight. The black dashed line shows the non-weighted radial trend of the template stacking also shown in Fig.~\ref{fig:ngc3059_stacking}. \textit{Bottom:} Deviation, in per cent, of the stacked radial trend from the template profile. The hatched area denotes the regime, where the prior, i.e. SNR/1, is detected in less than 20\,\% of the pixels.}
    \label{fig:ngc3059_weighted_stacking}
\end{figure}

The benefit of the above-described methodology is the potential recovery of faint emission while conserving the flux in each bin.
However, the drawback is that we might stack a few highly significant spectra with many noisy spectra, eventually leading to non-detection in the stacked spectra.
To overcome this, we can go beyond the ``equal weight per spectrum'' stacking described above, and weigh the spectra such that we obtain statistically more significant stacked spectra.\footnote{Though keep in mind that this weighted stacking scheme is in general not flux-conserving as opposed to the unweighted stacking introduced before.}
We compute the weighted stacks by multiplying the spectra ($T_{n,i}(v)$) with the associated weights ($w_i$) within each bin $n$.
Then, we sum up the weighted spectra and divide them with the sum of the weights:
\begin{align}
    T_{n,\mathrm{stack}}(v) = \dfrac{\sum\limits_{i=0}^{N_{\rm det}} T_{n,i}(v) \cdot w_i}{\sum\limits_{i=0}^{N_{\rm det}} w_i} \; .
    \label{EQU:weighted_stacking}
\end{align}
A useful weighting quantity could be the S/N or the line intensity of the prior.
Here, we showcase the latter, adopting an intensity-weighted stacking.
Thus, we obtain the radial trend of the prior-bright (e.g. CO-bright) regions.
We applied the intensity-weighted stacking to the above-introduced simulated observations analogous to the non-weighted stacking.
The radial stacking results are presented in Fig.~\ref{fig:ngc3059_weighted_stacking}.
We find that the stacked line intensities of the simulated cubes, excluding SNR/100, are consistent within 20\% with the true (weighted) trend.
Comparing with the un-weighted stacking (Fig.s \ref{fig:ngc3059_stacking} and \ref{fig:ngc3059_sigma_clipping}), we find better agreement and no negative bias at low detection fraction of the prior.
Thus, weighted stacking can indeed recover faint emission at larger galactocentric radii.
However, we note that weighted stacking does not conserve flux and must be interpreted with care, in particular when comparing to stacks derived with another, e.g. non-weighted, method.
Moreover, since, by construction, the intensity-weighted stacking used here is computing the weighted average stack over the detected pixels only, i.e. $N_\mathrm{det}$, the measured upper limits in the outer bins are much larger than what is obtained in the unweighted case.

\subsection{Stacking vs Averaging Integrated Intensities} 
\label{sec:stacking_vs_avg_ii}

\begin{figure}
    \centering
    \includegraphics[width=\columnwidth]{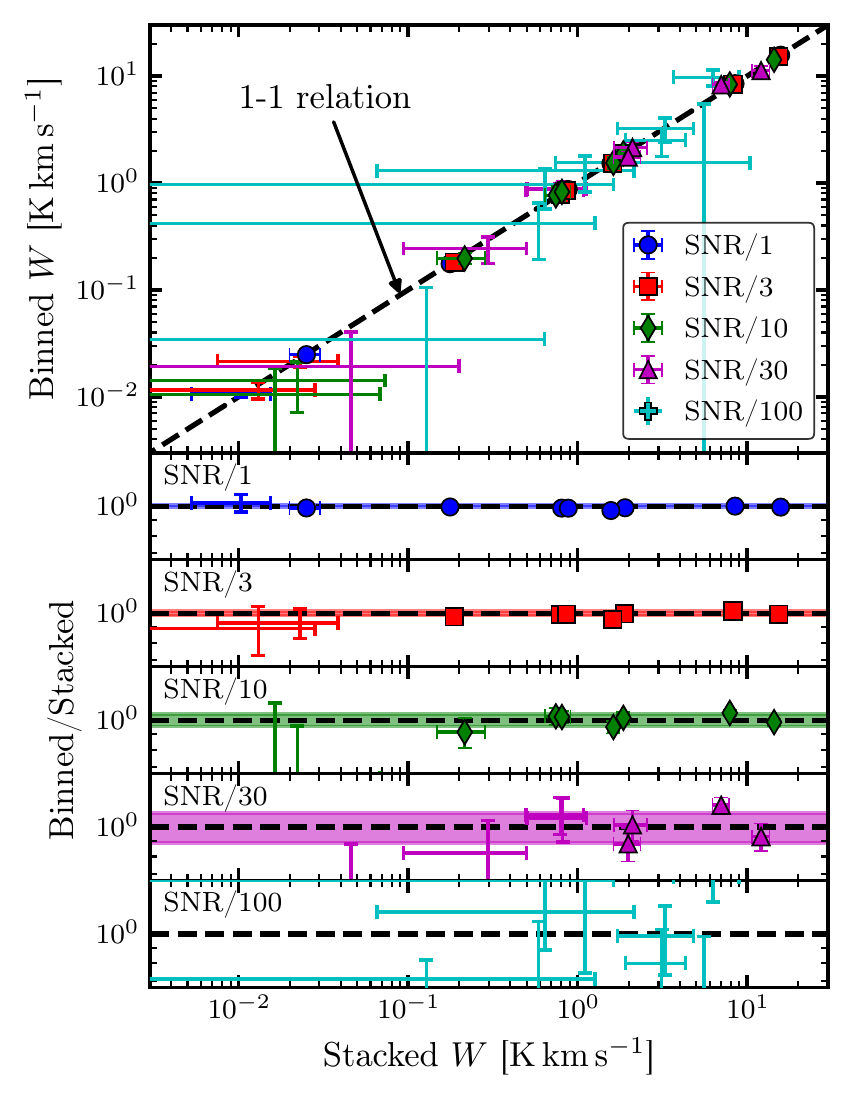}
     \caption{Binning vs. stacking. \textit{Top:} Binned means vs stacked integrated intensities in matched radial bins. The dashed line marks the 1-to-1 relation. Different markers indicate the values recovered from the different S/N cubes. Data above $3\sigma_W$ are shown as markers, else only the error bars are plotted. \textit{Bottom panels:} Ratio between binned mean and stacks against the stacked integrated intensities, separately for each SNR/X cube ($X={1, 3, 10, 30}$). The shaded areas indicate the respective 1-sigma scatter of the $3\sigma_W$ data.}
    \label{fig:ngc3059_stacking_vs_binning}
\end{figure}

Instead of averaging stacked spectra, it can be more convenient to average the integrated intensities within the same region/bin. With this approach, typically referred to as ``binning'', the main distinction to the stacking method is that we do not align the velocity field using a prior. Instead, we take advantage of the prior to create velocity masks for each individual spectrum, i.e. for each line of sight, which defines the velocity range over which each spectrum is integrated. The result is an integrated intensity (moment-0) map using a prior inferred velocity (field) mask \citep[see e.g.][for details about the masking process]{Gallagher2018a, Beslic2021, denBrok2022, Neumann2023}. Afterwards, we average the integrated intensities inside a given bin.

We apply the above averaging approach to the `12m+7m+tp' data sets at different noise levels using the same radial bins and compare the resulting average line intensities with the stacked line intensities computed as in Sec.~\ref{sec:stacking}. We find that both approaches lead to very similar average line intensities inside a given bin, without bias and small scatter of \{1, 2, 5, 10\}\% considering the significant measurements ($\mathrm{S/N}\geq 3$) of the SNR/1, 3, 10, 30 cubes (Fig.~\ref{fig:ngc3059_stacking_vs_binning}). In agreement with \citet{Gallagher2018b}, we conclude that spectral stacking and averaging masked moment-0 maps yield the same results within 10\%.
However, note that spectral stacking still offers the great advantage of also recovering mean line shape and thus mean kinematic information, which can not be obtained from the averaged integrated intensities.

\section{Conclusions} 
\label{sec:conclusions}

We performed spectral stacking of simulated interferometric data of the galaxy NGC\,3059 as a function of galactocentric radius at different noise levels and combining different telescope arrays.
Our main results are the following:
\begin{enumerate}
    
    \item Spectral stacking is able to recover the integrated intensities across most of the molecular disc where the prior is predominantly detected. In the most extreme case, we detect a stacked spectrum even in bins where the integrated intensities of the moment-0 map have a median S/N of 0.54. For this specific galaxy, all data above $3\sigma$ and $10\sigma$ agrees within 23\% and 15\%, respectively, with the expected values if the prior is detected in at least 36\% of the spectra contributing to the stack, i.e. within the inner 6\,kpc.

    \item  Using interferometric data only, i.e. without total power information, can filter out up to 30\% of the emission even at the typical PHANGS--ALMA sensitivity and even if the prior is predominantly significant. Even more extreme, for 10 times fainter lines, e.g. HCN(1--0), the 12m or 7m only configurations miss $\sim 50\%$ of the emission in the stacked spectra throughout the full molecular gas disc.

    \item The critical limitation of the spectral stacking method is connected to the quality of the prior used to align the velocity field and potentially the imaging procedure. If the prior is not detected across most of the bin, we expect to find systematically too low stacked line intensities. This might be improved by using low-resolution priors (e.g. \textsc{Hi} 21-cm line) or model priors, which provide a completely defined velocity field. However, we show that the discrepancy can also arise from the imaging of the interferometric data, e.g. if the deconvolution is not able to extract faint emission.
    
\end{enumerate}

This provides a concrete proof of concept that the stacking method works using combined interferometric and total power data on extended sources.
A key result of this analysis is that, at the typical PHANGS-ALMA setup, the spectral stacking method is able to recover the average integrated intensities within $\sim 23\%$ accuracy, if the prior is detected in at least $\sim 36\%$ of the bin's spectra. We also show that the noise estimated from the line-free parts of the stacked spectra captures the uncertainties of the line intensities with little bias (on average 10\% biased high) such that $3\sigma$ data can confidently be considered significant detection.

\begin{acknowledgements}
We would like to thank the anonymous referee for their
insightful comments that helped improve the quality of the paper.
LN acknowledges funding from the Deutsche Forschungsgemeinschaft (DFG, German Research Foundation) - 516405419.
AKL gratefully acknowledges support by grants 1653300 and 2205628 from the National Science Foundation, by award JWST-GO-02107.009-A, and by a Humboldt Research Award from the Alexander von Humboldt Foundation.
CE acknowledges funding from the Deutsche Forschungsgemeinschaft (DFG) Sachbeihilfe, grant number BI1546/3-1.
JP acknowledges support from the Programme National “Physique et Chimie du Milieu Interstellaire” (PCMI) of CNRS/INSU with INC/INP co-funded by CEA and CNES.
ER acknowledges the support of the Natural Sciences and Engineering Research Council of Canada (NSERC), funding reference number RGPIN-2022-03499.
ES acknowledges funding from the European Research Council (ERC) under the European Union’s Horizon 2020 research and innovation programme (grant agreement No. 694343).
\end{acknowledgements}

%
%

\bibliographystyle{aa} 
\bibliography{references.bib} 


\begin{appendix} 

\section{Additional material}
\label{sec:appendix}

\input{Tables/table_prior_detection.tex}

\begin{figure}
    \centering
    \includegraphics[width=0.95\columnwidth]{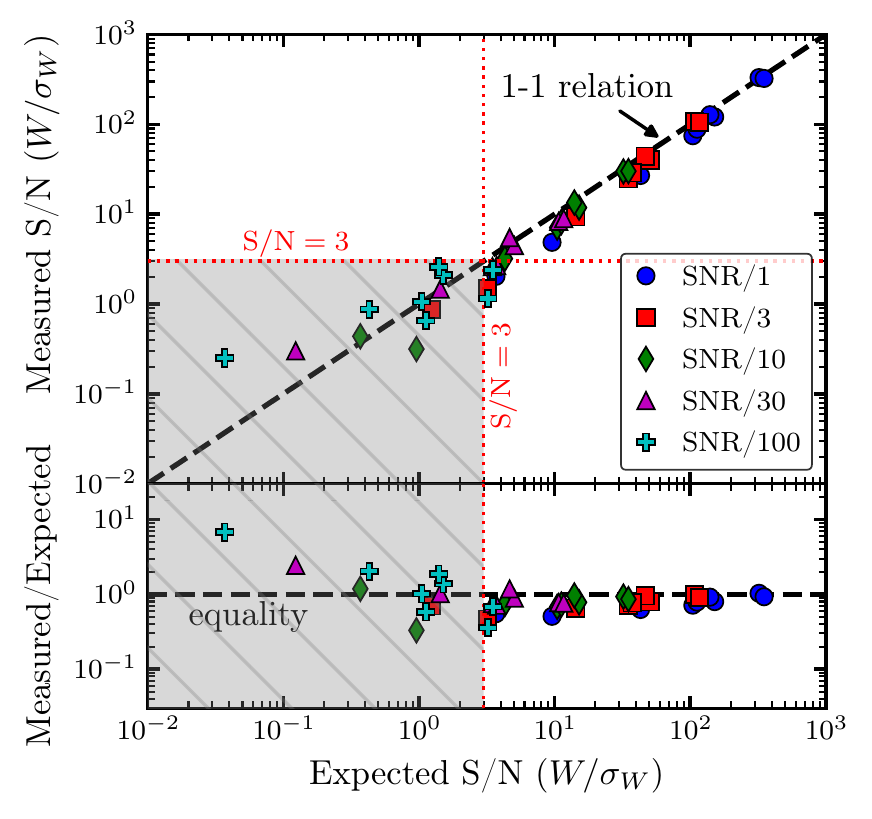}
     \caption{Measured vs. expected signal-to-noise ratio. \textit{Top:} Comparison between measured (12m+7m+tp) and expected signal-to-noise ratio of the integrated stacks. \textit{Bottom:} Ratio between the measured and the expected S/N against the expected uncertainties.}
    \label{fig:ngc3059_snr}
\end{figure}

\begin{figure}
    \centering
    \includegraphics[width=0.95\columnwidth]{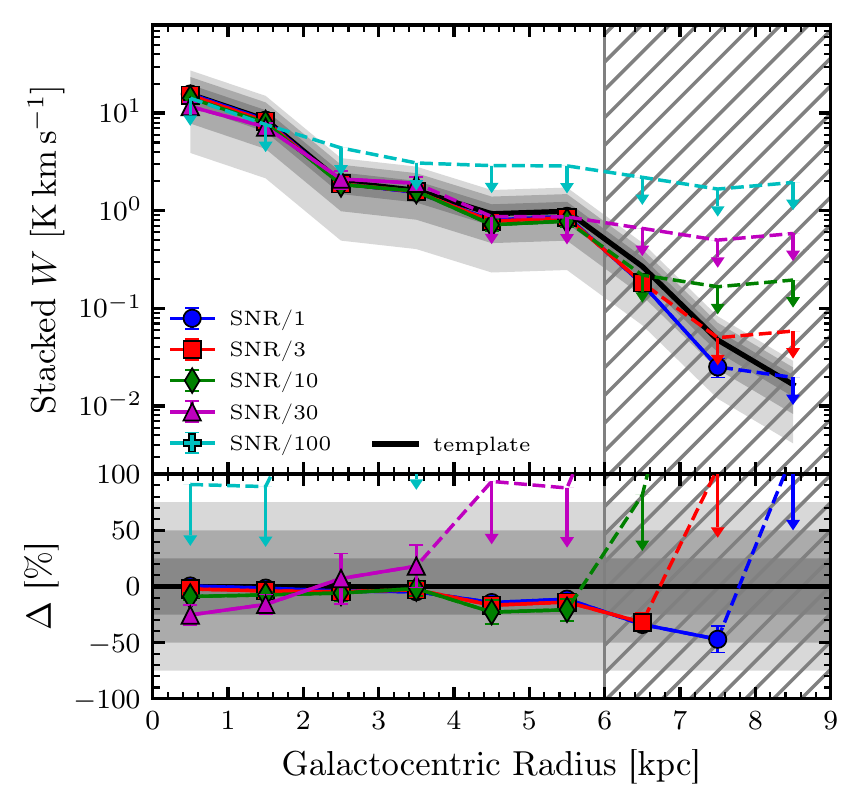}
     \caption{Template velocity field. Radial stacking similar to Fig.~\ref{fig:ngc3059_stacking} but using the template, i.e. the true intensity distribution, as prior to aligning the velocity field.}
    \label{fig:ngc3059_template_prior}
\end{figure}

\begin{figure}
    \centering
    \includegraphics[width=0.95\columnwidth]{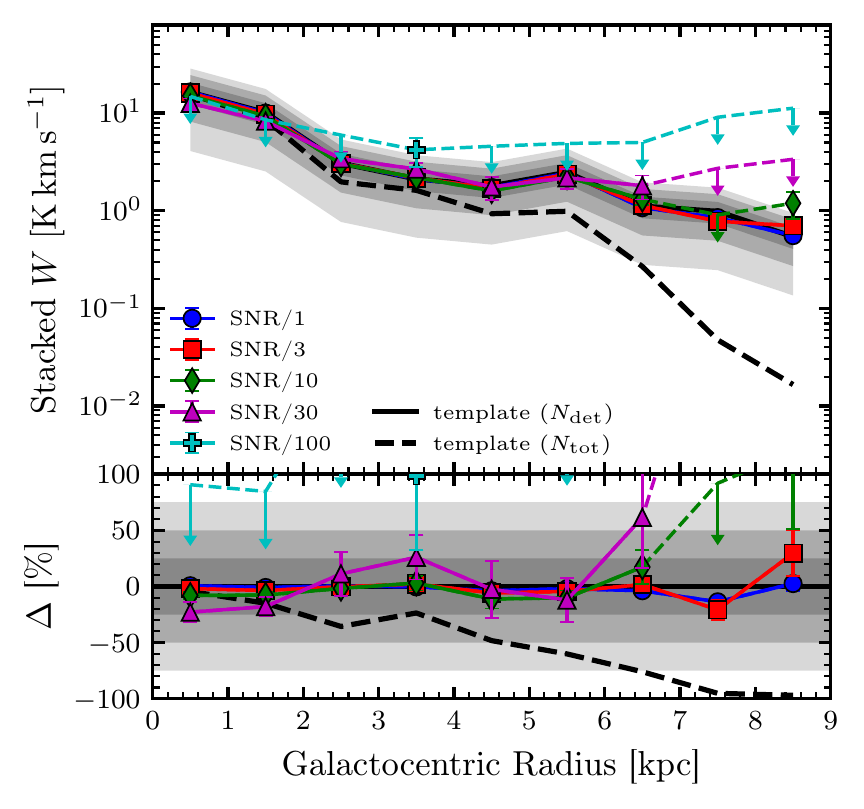}
     \caption{Average spectrum over prior-detected pixels. Radial stacking similar to Fig.~\ref{fig:ngc3059_stacking} but computing the average over the prior-detected spectra only as given by Eq.~\ref{equ:stacked_W_ignore_empties}.}
    \label{fig:ngc3059_ignore_empties}
\end{figure}

Table~\ref{tab:prior_detection} lists the detection fraction of the prior $F_\mathrm{det}=N_\mathrm{det}/N_\mathrm{tot}$ in each radial bin, where $N_\mathrm{det}$ is the number of prior-detected pixels and $N_\mathrm{tot}$ is the total number of pixels in that bin, respectively.

In Fig.~\ref{fig:ngc3059_snr}, we compare the S/N measured from the stacked spectra and the expected S/N that is inferred from the SNR/100 cube considered as a pure noise cube as described in Sec.~\ref{sec:uncertainties}.

In Fig.~\ref{fig:ngc3059_template_prior}, we show the spectral stacking as a function of galactocentric radius similar to Sec.~\ref{sec:W_recovery} but using the template data cube as prior instead of the SNR/1 cube. In this case, we have perfect alignment of the velocity field and are not limited by the significance of the prior.

In Fig.~\ref{fig:ngc3059_ignore_empties}, we show the radial trend obtained by taking the average spectrum over the prior-detected spectra only, i.e. by dividing the summed spectra by $N_\mathrm{det}$ instead of $N_\mathrm{tot}$ (Section~\ref{sec:stacking}). Eq.~\ref{equ:stacked_W} then changes to:
\begin{align}
    T_{n,\mathrm{stack}}(v) = \dfrac{1}{N_{\rm det}(n)} \sum\limits_{i=0}^{N_{\rm det}(n)} T_{n,i}(v) \;.
    \label{equ:stacked_W_ignore_empties}
\end{align} 
We find that the recovered stacks, computed from the prior-detected pixels, agree very well and without significant bias with the expected values measured in the prior-detected pixels (indicated by the solid line in Fig.~\ref{fig:ngc3059_ignore_empties}). However, using this method does not recover the true mean radial trend (dashed line), but, by construction, only considers the pixels, where the prior is detected and is thus biased high, especially at radii, where the prior detection fraction is low.

\end{appendix}

\end{document}

%% file: Tables/table_prior_detection.tex
\begin{table}
\caption{\label{t7}Prior detection fraction per radial bin.}
\centering
 \begin{tabular}{cccc}
    \hline\hline
    $r_{\rm gal}$ [kpc] &  $N_{\rm tot}$ &  $N_{\rm det}$ & $F_{\rm det}$ [\%] \\
    (1) & (2) & (3) & (4) \\
    \hline
    $0 - 1$ &     23 &       22 &   95.7 \\
    $1 - 2$ &     72 &       61 &   84.7 \\
    $2 - 3$ &    128 &       79 &   61.7 \\
    $3 - 4$ &    176 &      131 &   74.4 \\
    $4 - 5$ &    230 &      107 &   46.5 \\
    $5 - 6$ &    278 &      101 &   36.3 \\
    $6 - 7$ &    328 &       54 &   16.5 \\
    $7 - 8$ &    372 &       11 &    3.0 \\
    $8 - 9$ &    322 &        6 &    1.9 \\
    \hline\hline
 \end{tabular}
 \label{tab:prior_detection}
\tablefoot{(1) Radial bins, as illustrated in Figure~\ref{fig:ngc3059_stacking}. (2) Total number of spectra, i.e. pixels in moment-0 map, inside the respective bin. (3) Number of spectra, where the prior, i.e. ``SNR/1'', has been detected thus allowing velocity shuffling and spectral stacking. (4) Fraction of spectra used for stacking.}
\end{table}